\pdfoutput=1
\documentclass[
    aps,prb,twocolumn,
	groupedaddress,superscriptaddress,
	amsfonts,amssymb,amsmath,
	citeautoscript,longbibliography,
	letterpaper, nofootinbib
	]{revtex4-2}

\usepackage[utf8]{inputenc}
\usepackage[english]{babel}
\usepackage{lipsum}

\usepackage{microtype} 
\usepackage{xspace} 

\usepackage{txfonts}  
\usepackage{txfontsb} 

\usepackage{bm} 

\usepackage{xcolor}
\usepackage[]{graphicx} 
\graphicspath{{figs/}}

\usepackage[]{booktabs}
\usepackage{array}
\usepackage{layouts}
\usepackage{multirow}

\usepackage{enumerate}
\usepackage[inline]{enumitem}

\usepackage{xr}
\makeatletter
\newcommand*{\addFileDependency}[1]{
  \typeout{(#1)}
  \@addtofilelist{#1}
  \IfFileExists{#1}{}{\typeout{No file #1.}}
}
\makeatother

\usepackage{hyperref}
\hypersetup{colorlinks,
	linkcolor={blue!75!black!80!yellow},
	citecolor={blue!75!black!80!yellow},
	urlcolor={blue!75!black!80!yellow}
}

\usepackage[capitalize,nameinlink]{cleveref}

\crefname{subequations}{Eqs.}{Eqs.} 
\Crefname{subequations}{Eqs.}{Eqs.}
\crefformat{subequations}{#2Eqs.~(#1)#3}
\Crefformat{subequations}{#2Eqs.~(#1)#3}
\crefname{page}{p.}{p.} 
\crefname{table}{Table}{Tables}
\crefname{figure}{Figure}{Figures}
\crefname{section}{Section}{Sections}

\usepackage{placeins}

\usepackage{siunitx}
\DeclareSIUnit[number-unit-product = ]\percent{\char`\%} 

\usepackage[centering,hmargin=18mm,tmargin=29.4mm,bmargin=24mm]{geometry}

\usepackage{soul}

\usepackage{textcomp} 
\usepackage{xifthen}
\usepackage{etoolbox}
\newboolean{togglecomments}
\newboolean{toggletodos}
\newboolean{togglechanges}

\setboolean{togglecomments}{true}
\setboolean{toggletodos}{true}
\setboolean{togglechanges}{false} 

\newcommand{\textblacksquare}{$\blacksquare$}
\newcommand{\todo}[1]{\ifbool{toggletodos}%
	{\textcolor{green!60!black}{\small\textsf{{}\textsuperscript{\textsc{\textsf{todo}}}}[\ignorespaces#1]}} 
	{}}     
\newcommand{\comment}[2]{\ifbool{togglecomments}%
		{\textcolor{blue!70!black}{\small\sf\textsuperscript{\textsc{\textsf{\ignorespaces#1}}}[\ignorespaces#2]}} 
		{}}     
\newcommand{\swap}[2]{\ifbool{togglechanges}
	{\ignorespaces#2}  
	{\textcolor{red!70!black}{[\ignorespaces#1]}\textrightarrow{}\textcolor{green!50!black}{[\ignorespaces#2]}}}
\newcommand{\remove}[1]{\ifbool{togglechanges}
	{}    
	{\textcolor{blue}{\ignorespaces#1}}}
\newcommand{\inset}[1]{\ifbool{togglechanges}
	{\ignorespaces#1}  
	{\textcolor{green!50!black}{\ignorespaces#1}}}

\newcommand{\citeremind}[1]{%
	[\textcolor{blue!75!black!80!yellow}{\textblacksquare%
		\ifthenelse{\isempty{#1}}{}{\textsuperscript{\tiny\textsf{\ignorespaces#1}}}%
	}]\xspace}

\newcommand{\iu}{\mathrm{i}}

\newcommand{\ie}{i.e.,\@\xspace} 

\newcommand{\eg}{e.g.,\@\xspace}

\newcommand{\appropto}{\mathrel{\vcenter{
			\offinterlineskip\halign{\hfil$##$\cr
				\propto\cr\noalign{\kern.2pt}\sim\cr\noalign{\kern-2.5pt}}}}}

\DeclareFontFamily{U}{mathx}{\hyphenchar\font45}
\DeclareFontShape{U}{mathx}{m}{n}{<5> <6> <7> <8> <9> <10>
                                  <10.95> <12> <14.4> <17.28> <20.74> <24.88>
                                  mathx10}{}
\DeclareSymbolFont{mathx}{U}{mathx}{m}{n}
\DeclareFontSubstitution{U}{mathx}{m}{n}

\makeatletter
\newcommand{\raisemath}[1]{\mathpalette{\raisem@th{#1}}}
\newcommand{\raisem@th}[3]{\raisebox{#1}{$#2#3$}}
\makeatother

\renewcommand{\paragraph}[1]{\vskip 1ex\noindent\textbf{#1.}~}

\usepackage{braket}
\usepackage[eulergreek]{sansmath}
\makeatletter
\renewcommand\@make@capt@title[2]{%
    \@ifx@empty\float@link{\@firstofone}{\expandafter\href\expandafter{\float@link}}%
    \sisetup{math-sf=\textsf}%
    \sansmath\sffamily\textbf{#1\@caption@fignum@sep}#2 
}%

\makeatother

\setboolean{togglecomments}{true}
\setboolean{toggletodos}{true}
\setboolean{togglechanges}{false} 

\begin{document}
\title{Quantized crystalline-electromagnetic responses in insulators}

\author{Sachin Vaidya}
\email{svaidya1@mit.edu}
\thanks{S. V. and A. G. F. contributed equally}
\affiliation{Department of Physics, Massachusetts Institute of Technology, Cambridge, Massachusetts 02139, USA}
\author{Andr\'e Grossi Fonseca}
\email{agfons@mit.edu}
\affiliation{Department of Physics, Massachusetts Institute of Technology, Cambridge, Massachusetts 02139, USA}
\author{Mark R. Hirsbrunner}
\affiliation{Department of Physics and Institute for Condensed Matter Theory,
University of Illinois at Urbana--Champaign, Urbana,
IL 61801, USA}
\author{Taylor L. Hughes}
\affiliation{Department of Physics and Institute for Condensed Matter Theory,
University of Illinois at Urbana--Champaign, Urbana,
IL 61801, USA}
\author{Marin Solja\v ci\'c}
\affiliation{Department of Physics, Massachusetts Institute of Technology, Cambridge, Massachusetts 02139, USA}

\begin{abstract}
We introduce new classes of gapped topological phases characterized by quantized crystalline-electromagnetic responses, termed ``multipolar Chern insulators". 
These systems are characterized by nonsymmorphic momentum-space symmetries and mirror symmetries, leading to quantization of momentum-weighted Berry curvature multipole moments. 
We construct lattice models for such phases and numerically confirm their quantized responses. 
These systems exhibit bound charge and momentum densities at lattice and magnetic defects, and currents induced by electric or time-varying strain fields.
Our work extends the classification of topological matter by uncovering novel symmetry-protected topological phases with quantized responses.
\end{abstract}
\maketitle 

\paragraph{Introduction} 
Topology has become an indispensable framework for classifying and understanding phases of matter. 
A defining feature of topological phases is their quantized response to external probes. 
For instance, in the prototypical quantum Hall effect, an applied electric field induces a transverse Hall current~\cite{Laughlin1981, Thouless1982}, while magnetic flux defects bind electronic charge~\cite{Streda}.
These properties are governed by the Chern number, an integer-valued topological invariant corresponding to the Berry curvature integrated over the Brillouin zone. 
Extending beyond the conventional bulk-boundary correspondence, this framework of linear responses links electronic topology to response functions such as thermal transport~\cite{Luttinger1964, Kapustin2020}, geometric effects~\cite{Hughes2011, Teo2017, Gioia2021, Wang2021}, and higher-order electric multipoles~\cite{Benalcazar2017}, providing a unifying perspective across a variety of systems, including those with strong interactions.

Recent investigations have expanded this framework by considering the interplay of discrete translational symmetry and associated lattice defects in topological phases~\cite{Ran2009, Hughes2011, Gioia2021, Volovik2021, Hirsbrunner2024, Song2021, Pikulin2016, Grushin2016, Manjunath2021, Thorngren2018}.
Such defects, including strain fields and dislocations, serve as useful probes of electronic topology, as seen in weak topological insulators and semimetals.
These efforts have uncovered a hierarchy of crystalline-electromagnetic responses, characterized by higher moments of the Berry curvature~\cite{Mark2024}. 
Despite their richness, these responses have not been shown to exhibit quantization akin to the Chern number, raising the question of whether such quantization is possible and, if so, under what specific conditions it might arise.

In this work, we show that crystalline-electromagnetic responses are quantized through momentum-space nonsymmorphic symmetries. 
These symmetries, which arise from the interplay between gauge fields and point group operations, have recently garnered interest due to their connections with nonorientable Brillouin zones~\cite{KB1, grossi2023weyl, KB2, KB_acoustic1, KB_acoustic2, KB_acoustic3, KB_acoustic4, RP1, RP2, RP3, RP4, Konig2025, Rui2025} and projective symmetry algebras~\cite{Zhang2023}.
They have also been found to appear in certain moiré crystals~\cite{Calugaru2024} and spin space groups~\cite{Xiao2024}.
By imposing such symmetries, we identify new classes of insulating topological phases, termed ``multipolar Chern insulators," characterized by quantized momentum-weighted multipole moments of the Berry curvature. 
We show that these phases exhibit distinct responses to electromagnetic and translational gauge fields: magnetic and crystalline defects bind charge and momentum, while time-dependent lattice deformations and electric fields induce transverse currents. 
To illustrate the physics of these phases, we construct concrete lattice models for dipolar and quadrupolar Chern insulators and compute their quantized responses. 

\begin{figure*}
    \centering
    \includegraphics[scale=1]
    {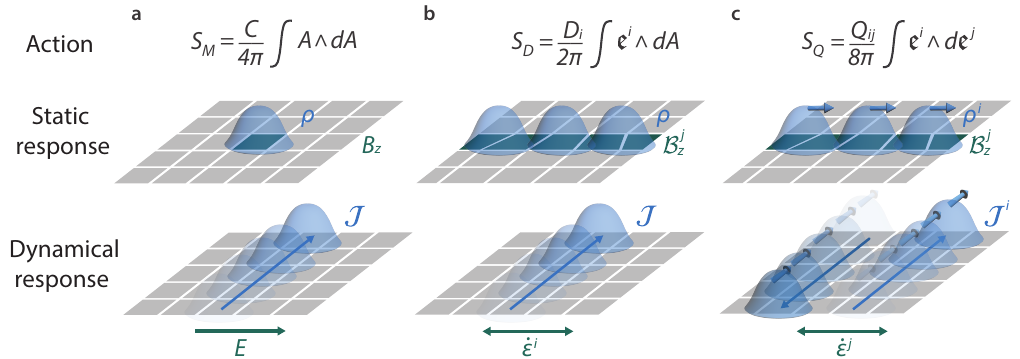}
    \caption{
    Hierarchy of electromagnetic and translational responses in topological phases.
    (a)~A Chern insulator is described by the usual Chern--Simons term (\cref{eq:Usual_CS}), with coefficient proportional to the Berry curvature monopole moment, \ie the Chern number.
    This action predicts charge density bound to magnetic flux defects and charge currents under electric fields.
    (b)~A dipolar Chern insulator is described by a mixed Chern--Simons term (\cref{eq:multipolar_actions}), with coefficient proportional to the Berry curvature dipole moment.
    This action predicts charge density bound to translation flux defects, such as strain, and charge currents under time-varying strain fields.
    (c)~A quadrupolar Chern insulator is described by a purely crystalline Chern--Simons term (\cref{eq:multipolar_actions}), with coefficient proportional to the Berry curvature quadrupole moment.
    This action predicts momentum density (represented by the arrows attached to the charge density) bound to translation flux defects and momentum currents under time-varying strain fields. In the latter case, we note that the absence of a net charge current implies the presence of counter-propagating charge currents, each carrying momentum in the same direction.
        }
    \label{fig:figure1}
\end{figure*}

\paragraph{Crystalline-electromagnetic responses}
We motivate the discussion of responses associated with discrete translational symmetry by drawing parallels between electromagnetic and translational defects.
Consider first a charged particle on a lattice pierced by a localized magnetic flux through a plaquette, \ie a magnetic flux defect. 
When a particle traverses a closed path around this defect, its wavefunction acquires an Aharonov--Bohm phase proportional to the particle's charge and the enclosed flux. 
This phase corresponds to the line integral of the electromagnetic gauge field, $A_\mu$, along the path. 
Similarly, a particle traversing a closed path encircling a dislocation, \ie a translation defect, experiences a transformation of its wavefunction. 
However, in this case, the particle is translated by the Burgers vector of the dislocation due to the elastic deformation of the lattice~\cite{Teo2017}. 

This perspective motivates the introduction of a set of translational gauge fields, $\mathfrak{e}_\mu^a$, where $a$ runs over the spatial directions, which are associated to translational symmetry in the $a$-th direction.
These gauge fields couple to electronic degrees of freedom and encapsulate general lattice deformations and dislocation defects. Specifically, the line integral of $\mathfrak{e}_\mu^a$ around a closed loop corresponds to the $a$-th component of the net Burgers vector of all dislocations enclosed within the region.
In addition, integrating $\mathfrak{e}_\mu^a$ on paths encircling the entire lattice captures strains and shears. 
By introducing the gauge fields $\mathfrak{e}_\mu^a$, electromagnetic and translational defects can be described on equal footing.
However, care must be taken with the definition of $\mathfrak{e}_\mu^a$, as it arises from gauging discrete translational symmetry~\cite{Manjunath2021, Thorngren2018, Song2021}.

With this context in mind, we now analyze low-energy topological responses in two dimensions involving both electromagnetic and translation gauge fields. 
A Chern insulator is described by the purely electromagnetic Chern--Simons action
\begin{equation}
\label{eq:Usual_CS}
S_M = \frac{C}{4\pi}\int \mathrm{d}^2 \mathbf{r} \,\mathrm{d} t\,\varepsilon^{\mu\nu\rho} A_\mu \partial_\nu A_\rho \equiv \frac{C}{4\pi}\int A \wedge dA,
\end{equation}
where $C$ is the Chern number, Greek letters index spacetime dimensions, and $\varepsilon^{\mu\nu\rho}$ is the three-dimensional Levi--Civita symbol. 
The topological response is derived by varying $S_M$ with respect to $A_\mu$ to obtain the current $\mathcal{J}^\mu$. 
For spatial components, this procedure yields the dynamical response of transverse charge currents flowing under applied electric fields, \ie the quantum Hall effect. 
On the other hand, varying with respect to the temporal component yields the St\v reda formula~\cite{Streda}
\begin{equation}
    \rho = \frac{C}{2\pi} B_z,
\end{equation}
indicating that charge density, $\rho,$ is locally bound to magnetic flux defects generated by the perpendicular magnetic field $B_z$ (\cref{fig:figure1}a).

In a similar vein, one can construct actions involving both $A_\mu$ and $\mathfrak{e}^a_\mu$, as well as those defined exclusively in terms of $\mathfrak{e}^a_\mu$. 
Whereas the former describe mixed crystalline-electromagnetic responses, the latter are purely crystalline. 
The simplest of these actions are of the Chern--Simons form~\cite{Mark2024}:
\begin{equation}
\label{eq:multipolar_actions}
    S_D = \frac{D_i}{2\pi}\int\mathfrak{e}^i \wedge dA, \quad 
    S_Q = \frac{Q_{ij}}{8\pi}\int\mathfrak{e}^i \wedge d\mathfrak{e}^j,
\end{equation}
where throughout we set $e = \hbar = 1$, and
coefficients $D_i, Q_{ij}$ to be discussed below.
The mixed term $S_D$, when varied with respect to $A_0$ or $\mathfrak{e}^a_0$, predicts charge density bound to lattice defects or momentum density bound to magnetic flux defects, respectively. 
Similarly, varying $S_D$ with respect to $A_i$ and $\mathfrak{e}^a_i$ describes charge currents induced by time-dependent strains and momentum currents generated in response to electric fields~(\cref{fig:figure1}b). 
The purely crystalline term $S_Q$ governs momentum-like responses to crystalline deformations. 
Variation with respect to the time component describes momentum density bound to translationally symmetric lattice defects, such as homogeneous strain fields or dislocation lines. 
Varying with respect to the spatial components, on the other hand, predicts a momentum current arising from the time derivative of strain and shear deformations~(\cref{fig:figure1}c), bearing close similarity to the Hall viscosity~\cite{Avron1995, Read2009, Hughes2011, Hughes2013, Parrikar2014, Bradlyn2015, Rao2020, Fruchart2023}.

Recent studies have shown using dimensional reduction and augmentation techniques that these responses form a hierarchical structure reminiscent of the classical multipole expansion~\cite{Mark2024}. 
In this hierarchy, the coefficients in \cref{eq:multipolar_actions} correspond to momentum-weighted dipole ($\boldsymbol{D}$) and quadrupole ($\boldsymbol{Q}$) moments of the Berry curvature whose components are:
\begin{equation}
\begin{split}
\label{eq:BC_moments}
    D_i = \frac{1}{2\pi} \int_{BZ} d^2k\, k_i \,\mathcal{F}(\mathbf{k}), \quad 
    Q_{ij} = \frac{1}{\pi} \int_{BZ} d^2k\, k_i k_j \, \mathcal{F}(\mathbf{k}),
\end{split}
\end{equation}
where $\mathcal{F}(\mathbf{k})=\partial_{k_x} \mathcal{A}_y(\mathbf{k}) - \partial_{k_y} \mathcal{A}_x(\mathbf{k})$ is the Berry curvature, and $\mathcal{A}_i(\mathbf{k}) = \iu \braket{ u_\mathbf{k}|\partial_{k_i} u_\mathbf{k}}$ the Berry connection.~\footnote{We note that these multipole moments of the Berry curvature are distinct from those discussed in the context of the quantum nonlinear Hall effect \cite{sodemann2015quantum, zhang2023higher}, which involve momentum-space derivatives of the Berry curvature.}
Such coefficients are only well-defined if all lower-moments vanish, otherwise they are dependent on the choice of the Brillouin zone origin.
For example, the dipolar action $S_D$ is well-defined for time-reversal symmetric insulators, which, by symmetry, would have a vanishing monopole term, \ie vanishing Chern number. 
Finally, we can also use these expressions to evaluate discrete changes in $D_i$ and $Q_{ij}$ as the system is driven through a gapless phase with Dirac points.
Using the fact that each Dirac point sources $\pi$ Berry curvature, the changes evaluate to
\begin{equation}
\begin{split}
    \Delta D_i = \sum_a \chi_a  k^a_i , \quad
    \Delta Q_{ij} = 2 \sum_a \chi_a  k^a_i k^a_j,
\label{eq:BC_moments_jump}
\end{split}
\end{equation}
where the summation runs over all Dirac points in the Brillouin zone, $\chi^a$ is the helicity of the Dirac point, and the factor of $2$ comes from the sign flip of $\chi^a$ upon crossing the transition point. 
\cref{eq:BC_moments} and \cref{eq:BC_moments_jump} fully characterize the low-energy leading crystalline-electromagnetic responses in phases with vanishing Chern number.

We now turn to the quantization of the coefficients in \cref{eq:BC_moments}, which, unlike the Chern number, requires symmetries.
To understand how symmetries might enforce quantization, we begin by considering the atomic limit, in which these moments vanish.
Transitioning from this atomic limit to an insulating phase having quantized coefficients requires two essential conditions.
First, the Berry curvature moments must undergo a quantized jump at the gap-closing points that mediate the transition.
Thus, from \cref{eq:BC_moments_jump}, the Dirac points at the transition must be constrained to lie at specific high-symmetry locations in momentum space, fixing the momentum-space distance between them.
Second, once the gap reopens and the system enters an insulating phase, symmetries must further constrain the Berry curvature distribution to preclude deformations that shift the moments away from quantized values.
As we show in the next section, momentum-space nonsymmorphic symmetries, combined with mirror symmetries, are sufficient to enable this quantization.

\begin{figure}[t]
    \centering
    \includegraphics[scale=1]
    {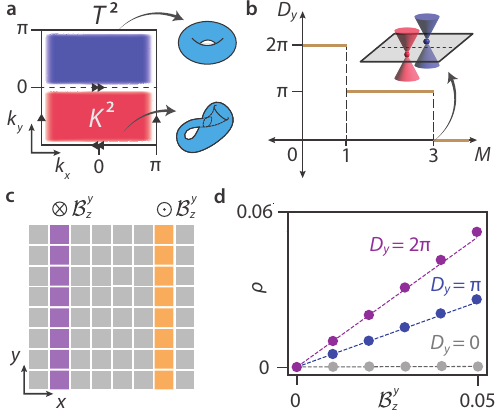}
    \caption{%
    (a)~Generic Berry curvature distribution across the 2D Brillouin zone for a system with symmetries in \cref{eq:BC_dipole_symms}, splitting the torus into two halves, each topologically equivalent to a Klein bottle.
    Boundary identifications are indicated by arrows.
    (b)~Phase diagram for the $y$-component of the Berry curvature dipole moment vector, $D_y$, as a function of $M$ for the model in~\cref{eq:BC_dipole_model}. (c)~Configuration used for the response calculation: a system with translational magnetic fluxes $\mathcal{B}_z^y$ of opposite signs inserted through two lines of plaquettes.
    (d)~Linear response of the charge density bound to the translation flux for the topological (purple, blue) and trivial (gray) phases at $M=0$, $M=2$ and $M=4$ respectively. 
    The dashed lines corresponds to the theoretical prediction from \cref{eq:linear_resp_dipole}.
    }
    \label{fig:figure2}
\end{figure}

\paragraph{Dipolar Chern insulator}
We first focus on the dipolar Chern insulator---a gapped phase with vanishing Chern number and quantized Berry curvature dipole moment.
In order to quantize $D_i$, we impose both a momentum-space glide symmetry and $y$-mirror symmetry:
\begin{equation}
\label{eq:BC_dipole_symms}
    \begin{split}
        &\mathcal{G}_x: H(k_x, k_y) = U_1^{\dag} H(-k_x, k_y + \pi)U_1, \\
        &M_y: H(k_x, k_y) = U_2^{\dag} H(k_x, -k_y)U_2,
    \end{split}
\end{equation}
where $U_i$ are unitary operators, and we have set the lattice constant to unity.
The symmetry $\mathcal{G}_x$ partitions the Brillouin zone torus into two halves, each of which have the topology of a Klein bottle  $K^2$~\cite{KB1, grossi2023weyl} (\cref{fig:figure2}a).
The Berry curvature distribution in each half is related through $\mathcal{G}_x$ by a sign flip, ensuring a vanishing Chern number.
Furthermore, $M_y$ requires the Berry curvature to vanish along the Klein bottle boundaries and to spread symmetrically along $k_y$ in the gapped phase.
In the Supplemental Material (SM) we prove that  $\boldsymbol{D}$ is quantized under these symmetries, with components $D_i = \pi n, n \in \mathbb{Z}$.

We now explicitly demonstrate this quantization via a two-band, spinless model on the square lattice with a Bloch Hamiltonian of the form
\begin{align}
    H(\mathbf{k}) = \mathbf{d}\cdot \boldsymbol{\sigma} = d_x(\mathbf{k}) \sigma_x + d_y(\mathbf{k}) \sigma_y + d_z(\mathbf{k}) \sigma_z,
    \label{eq:Bloch_H}
\end{align}
where $\sigma_{x,y,z}$ are the Pauli matrices, and $H(\mathbf{k})$ obeys the symmetries in \cref{eq:BC_dipole_symms}.
In particular, we consider
\begin{align}
    \begin{split}
    \label{eq:BC_dipole_model}
    d_x(\mathbf{k}) &= \sin{k_x} \sin{k_y}, \\
    d_y(\mathbf{k}) &= \cos{k_x}\sin{2k_y}, \\
    d_z(\mathbf{k}) &= M - \cos{k_x} + 2\cos{2k_y},
    \end{split}
\end{align}
where $M$ is a tunable mass parameter, and we choose $U_1 = \mathbb{1}$ and $U_2 = \sigma_z$.
For both $-1 < M < 1$ and $1 < M < 3$, the system is a dipolar Chern insulator characterized by $D_x = 0, D_y = 2\pi$ and $D_x = 0, D_y = \pi$ respectively.
At $M = 3$, the gap closes at two Dirac points located at the $K^2$ centers $ (0, \pm \pi/2)$, and for $M>3$ the model transitions to a trivial phase where $D_x = D_y = 0$ (\cref{fig:figure2}b).
A model having an identical phase diagram, but with nonzero $D_x$ and vanishing $D_y,$ can be derived from \cref{eq:BC_dipole_model} by interchanging $k_x$ and $k_y$.
In the SM we discuss the observed boundary signatures of these phases and their connection to phases with sub-Brillouin zone topology~\cite{Chen2024}.

We now calculate the response of the non-trivial gapped phase. 
As discussed earlier, the dipolar action $S_D$ in \cref{eq:multipolar_actions} predicts a mixed crystalline-electromagnetic response in which translation magnetic fluxes, generated by strain or dislocations, bind charge density. 
To verify this, we consider a ribbon geometry having boundaries along the $x$-direction, and introduce two columns of dislocations near the right and left boundaries, ensuring translational symmetry along $y$~(\cref{fig:figure2}c). 
We choose each dislocation to have a Burgers vector pointing in the $y$-direction, such that the opposing columns of dislocations have opposite Burgers vectors and therefore generate opposite translational magnetic fluxes, $\mathcal{B}^y_z = \partial_x \mathfrak{e}^y_y - \partial_y \mathfrak{e}^y_x$. 
Hence, from our response we expect to find bound charge density that is opposite for each column.
Numerically, such columns of dislocations can be implemented by modifying the hopping terms in the region between the columns with generalized $k_y$-dependent Peierls factors $\exp(\iu k_y \mathcal{B}^y_z)$~\cite{Mark2024}.
This amounts to shifting the hopping amplitudes between the rows of dislocations by $k_y \rightarrow k_y (1 + \mathcal{B}^y_z)$.
Physically, this can be thought of as straining the lattice along $y$ in the region between the columns. 
In this geometry, varying the action $S_D$ with respect to $A_0$ yields a St\v reda-like formula
\begin{equation}
\label{eq:linear_resp_dipole}
    \rho = \frac{D_y}{2\pi}\,\mathcal{B}^y_z.
\end{equation}
By evaluating the charge density near one column of dislocations, we show that this linear relationship holds for the model in \cref{eq:BC_dipole_model} in the topological and trivial phases (\cref{fig:figure2}d), confirming that the slopes take quantized values proportional to $D_y$.

We now comment on the connection between electric polarization and the response coefficient $D_i$. 
As shown in Ref.~\cite{Mark2024}, the electric polarization, $P_i$, satisfies the relation $P_i = \varepsilon^{ij} D_j + (1/2\pi)W_i$, where $\varepsilon^{ij}$ is the two-dimensional Levi--Civita symbol and $W_i = \oint dk_i \mathcal{A}_i(k_i, k_j = \pi)$ is an additional Wilson loop contribution. 
Under inversion symmetry, polarization is quantized, which 
follows from the fact that $D_j = 0$ and $W_i$ is quantized to either 0 or $\pi$. 
In the presence of the symmetries in \cref{eq:BC_dipole_symms}, we find that the electric polarization is also quantized, now due to the quantization of $D_j$ and vanishing $W_i$.

Finally, we note several subtle distinctions between our work and prior approaches that equate the response coefficient of the dipolar action with the electric polarization~\cite{Song2021, Manjunath2021}. 
As derived in \cite{Mark2024} and stated above, the coefficient of the dipolar action is not, in general, equivalent to electric polarization, except in the presence of additional crystalline symmetries or in the semimetallic limit. 
This distinction becomes particularly relevant in the presence of lattice defects. 
We have found that the response to a dislocation implemented via a momentum-dependent Peierls phase—effectively modeling a uniform strain—differs from that of an explicit dislocation formed by the addition or removal of lattice sites, even though both are characterized by a Burgers vector. 
Numerical calculations in \cref{fig:figure2}d confirm that the former configuration yields bound charge consistent with our response coefficient, rather than with the conventional polarization formula. 
This highlights a fundamental difference between strain and discrete dislocations.

\begin{figure}[t]
    \centering
    \includegraphics[scale=1]
    {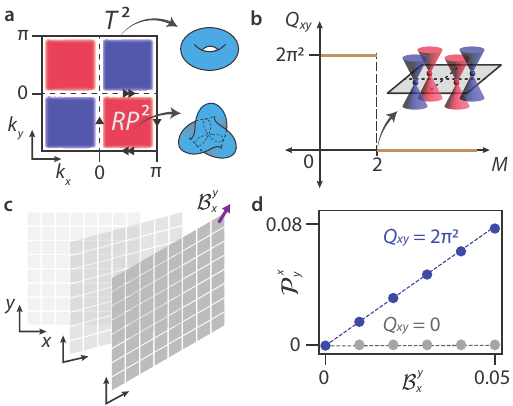}
    \caption{%
    (a)~Generic Berry curvature distribution across the 2D Brillouin zone for a system with symmetries in \cref{eq:BC_quadrupole_symms}, splitting the torus into four quarters, each topologically equivalent to a real projective plane.
    Boundary identifications are indicated by arrows.
    (b)~Phase diagram for the off-diagonal component of the Berry curvature quadrupole moment tensor, $Q_{xy}$, as a function of $M$ for the model in~\cref{eq:BC_quadrupole_model}.
    (c)~Configuration used for the response calculation: a homogeneously sheared lattice, for which electrons propagating along $x$ are translated in $y$, which corresponds to threading a constant translational magnetic flux $\mathcal{B}^y_x$.
    (d)~Linear response of the momentum polarization to a lattice shear for the topological (blue) and trivial (gray) phases at $M=1$ and $M=3$ respectively.
    }
    \label{fig:figure3}
\end{figure}

\paragraph{Quadrupolar Chern insulator}
We next turn our attention to the quadrupolar Chern insulator---a gapped phase characterized by a vanishing Chern number and vanishing Berry curvature dipole moment, and a quantized, non-vanishing Berry curvature quadrupole moment.
Such a phase can be realized \eg by enforcing three momentum-space nonsymmorphic symmetries:
\begin{equation}
\label{eq:BC_quadrupole_symms}
    \begin{split}
        &\overline{\mathcal{G}}_x: H(k_x, k_y) = U_1^{\dag}H(-k_x+\pi, k_y + \pi)U_1, \\
        &\overline{\mathcal{G}}_y: H(k_x, k_y) = U_2^{\dag}H(k_x + \pi, -k_y + \pi)U_2, \\
        &\mathcal{S}: H(k_x, k_y) = U_3^{\dag}H(k_y, -k_x+\pi)U_3,
    \end{split}
\end{equation}
where $U_i$ are unitary operators, we have once again set the lattice constant to unity, and overbars are used to distinguish these symmetries from those in \cref{eq:BC_dipole_symms}, which shift only one momentum direction.
The combination of $\overline{\mathcal{G}}_x$ and $\overline{\mathcal{G}}_y$ divides the Brillouin zone torus into four quadrants, each having the topology of the real projective plane, $RP^2$ (\cref{fig:figure3}a). 
The Berry curvature distribution within each quadrant is related by successive operations of $\overline{\mathcal{G}}_x$ and $\overline{\mathcal{G}}_y$, which ensures both that the Chern number and Berry curvature dipole moment vanish, and that the Berry curvature itself exactly vanishes along the $RP^2$ boundaries. 
The third symmetry $\mathcal{S}$ requires that the Berry curvature spreads along $k_x$ and $k_y$ with $C_4$ symmetry about the four $RP^2$ centers, \ie at $(\pm \pi/2, \pm \pi/2$), in the gapped phase.
In the SM  we prove that the Berry curvature quadupole moment, $\boldsymbol{Q}$, is quantized under these symmetries, with components $Q_{ij} \in 2\pi^2 n, n \in \mathbb{Z}$.

A concrete example of a two-band model hosting a quadrupolar Chern insulator phase has the form:
\begin{align}
    \begin{split}
    \label{eq:BC_quadrupole_model}
    d_x(\mathbf{k}) &= \sin{k_x} \sin{2k_y}, \\
    d_y(\mathbf{k}) &= \sin{k_y}\sin{2k_x}, \\
    d_z(\mathbf{k}) &= M + \cos{2k_x} + \cos{2k_y},
    \end{split}
\end{align}
where $M$ is a tunable mass parameter, and we choose $U_1 = U_2 = \mathbb{1}, U_3 = \exp(i\pi\sigma_z/2)$. For $0 < M < 2$, the system is a quadrupolar Chern insulator characterized by $Q_{xx} = Q_{yy} = 0, Q_{xy} = 2\pi^2$.
At $M = 2$, the gap closes  at four Dirac points located at the $RP^2$ centers, and for $M>2$  the system is in the trivial phase $Q_{ij} = 0$ (\cref{fig:figure3}b).
A model with an identical phase diagram but with $Q_{xx} = -Q_{yy} = 2\pi^2, Q_{xy} = 0$ can be derived from \cref{eq:BC_quadrupole_model} by rotating momentum space by $\pi/4$.
In the SM we discuss the observed boundary signatures of these phases.

To analyze the responses of this phase, we use the action $S_Q$ from \cref{eq:multipolar_actions}, which predicts purely crystalline responses where strain and dislocations bind momentum density instead of charge density. 
As an illustration, we focus on the dynamical response obtained by varying $S_Q$ with respect to $\mathfrak{e}_y^x$, which predicts a $k_x$ momentum current to flow along $y$ when threading a time-varying $\mathfrak{e}_x^y$ through the system:
\begin{equation}
\label{eq:quad_response}    \mathcal{J}_y^x=\frac{Q_{xy}}{4\pi} \partial_t \mathfrak{e}_x^y,
\end{equation}
where $\mathcal{J}_a^\mu = k_a\, \mathcal{J}^\mu$, with $\mathcal{J}^\mu$ the charge current along $\mu$.
Physically, a non-zero value of $\mathfrak{e}_x^y$ means that an electron traversing a path along $x$ is translated along the $y$-direction, \ie a shear deformation. In \cref{fig:figure3}c we depict a homogeneous shear deformation, a configuration that preserves translational symmetry along both directions. We employ this geometry to efficiently compute the response in momentum space.

We carry out a time-independent calculation of the flowing momentum current by evaluating the momentum polarization $\mathcal{P}_y^x(t)=\int_{0}^{t} dt'\, \mathcal{J}_y^x$. 
Integrating both sides in \cref{eq:quad_response} yields 
\begin{equation}
\label{eq:quad_response_TI}
\mathcal{P}_y^x(t) = \frac{Q_{xy}}{4\pi}\mathfrak{e}_x^y(t).
\end{equation}
The momentum polarization can be numerically calculated by a generalized surface charge theorem~\cite{vanderbilt1993electric}, $\mathcal{P}_y^x = \int d^2\mathbf{k} \, k_y \mathcal{A}_x(\mathbf{k})$.
Because a finite $\mathfrak{e}_x^y$ represents a shear where electrons moving in $x$ are translated in $y$, there is an additional phase/translation operator $\exp \left( \iu k_y \int dx\, \mathfrak{e}^y_x \right)$, which corresponds to a global translation flux $\mathcal{B}^y_x$ through a cycle of the position-space torus.
Therefore, a homogeneous shear can be numerically implemented via the substitution $k_x \rightarrow k_x + \mathcal{B}_x^y k_y$ in the Bloch Hamiltonian. We confirm that the linear relationship in \cref{eq:quad_response_TI} holds for the model in \cref{eq:BC_quadrupole_model} in both the topological and trivial phases (\cref{fig:figure3}d), confirming that the slopes take quantized values proportional to $Q_{xy}$.

\paragraph{Conclusions}
In this work, we have explored topological phases exhibiting quantized crystalline-electromagnetic responses, analyzed both through general theoretical considerations and concrete lattice models. 
The coefficients of these responses are determined by momentum-weighted multipole moments of the Berry curvature and are quantized in the presence of momentum-space nonsymmorphic symmetries and mirror symmetries. 
These effective actions describe physical phenomena such as bound charge and momentum densities at magnetic and lattice defects, as well as charge and momentum currents induced by electric fields and time-dependent strains, as explicitly demonstrated in our illustrative models.

More broadly, our work reveals a novel connection between nonorientable momentum-space manifolds and topological response theory. 
This motivates further exploration of the interplay between higher-order quantized crystalline-electromagnetic responses that emerge in systems with momentum-space nonsymmorphic symmetries in interacting systems, in higher dimensions, and in non-Hermitian systems~\cite{Konig2025, Rui2025}.
The models and responses presented here are readily accessible across various physical platforms~\cite{lin2023topological}, including electrical circuits~\cite{yamada2022bound}, ring-resonator lattices, acoustic crystals~\cite{ye2022topological}, mechanical systems~\cite{paulose2015topological}, photonic crystals~\cite{li2018topological, Barsukova2023}, and optical waveguides~\cite{lustig2022photonic, 3DPhCs}. 
Moreover, the recent predictions of momentum-space nonsymmorphic symmetries in both a moiré crystals~\cite{Calugaru2024} and in spin space groups~\cite{Xiao2024} offer an exciting prospect for observing these phenomena in solid-state systems.

\vskip 2ex
\paragraph{Acknowledgements} 
We thank Thomas Christensen, Mikael C. Rechtsman, Christopher Fechisin and Yuxuan Zhang for stimulating discussions.
A. G. F.\ acknowledges support from the Henry W. Kendall Fellowship and the Whiteman Fellowship, and thanks the University of S\~ao Paulo for its hospitality, where part of this work was completed.  
S. V., M. R. H., T. L. H., and M. S.\ acknowledge support from the U.S.\ Office of Naval Research (ONR) Multidisciplinary University Research Initiative (MURI) under Grant No.\ N00014-20-1-2325 on Robust Photonic Materials with Higher-Order Topological Protection.
This material is based upon work also supported in part by the U. S. Army Research Office through the Institute for Soldier Nanotechnologies at MIT, under Collaborative Agreement Number W911NF-23-2-0121, and also in part by the Air Force Office of Scientific Research under the award number FA9550-21-1-0299. 

\paragraph{Supplementary Material}
The supplementary material contains proofs for the quantization of response coefficients, a discussion on the bulk-boundary correspondence for these phases, and additional references~\cite{schindler2021noncompact, zhu2023z, alexandradinata2024quantization, nelson2022delicate}.

\bibliography{main}

\end{document}


\title{\texorpdfstring{
        SUPPLEMENTAL MATERIAL\\[1ex]
         Quantized electromagnetic-crystalline responses in insulators
        }
        {Supplemental Material}
       }

\author{Sachin Vaidya}
\email{svaidya1@mit.edu}
\affiliation{\mitaffil}

\author{Andr\'e Grossi Fonseca}
\email{agfons@mit.edu}
\affiliation{\mitaffil}

\author{Mark R. Hirsbrunner}
\affiliation{\uiucaffil}

\author{Taylor L.~Hughes}
\affiliation{\uiucaffil}

\author{Marin Solja\v{c}i\'{c}}
\affiliation{\mitaffil}

\maketitle

\setlength{\parindent}{0em}
\setlength{\parskip}{.5em}

\noindent{\small\textbf{\textsf{CONTENTS}}}\\ 
\twocolumngrid
\begingroup
    \let\bfseries\relax 
    \deactivateaddvspace 
    \deactivatetocsubsections 
    \makeatletter\@starttoc{toc}\makeatother 
\endgroup
\onecolumngrid

\count\footins = 1000 
\interfootnotelinepenalty=10000 

\section{Quantization of Berry curvature multipole moments}

In this section, we consider the action of various momentum-space nonsymmorphic symmetries and prove that they lead to quantized Berry curvature multipole moments.

\subsection{Berry curvature dipole moment}
We first consider systems with a momentum-space glide symmetry and a mirror symmetry.
The explicit action of these symmetries on the Hamiltonian is as follows:
\begin{equation}
\label{eq:BC_dipole_symms}
    \begin{split}
        &\mathcal{G}_x: H(k_x, k_y) = U_1^{\dag} H(-k_x, k_y + \pi)U_1, \\
        &M_y: H(k_x, k_y) = U_2^{\dag} H(k_x, -k_y)U_2, 
    \end{split}
\end{equation}
where $U_i$ are unitary operators.
We will show that these symmetries lead to the quantization of the $y$ component of the Berry curvature dipole moment, $D_{y} = \frac{1}{2\pi} \int_{\text{BZ}} \mathrm{d}^2\mathbf{k}\, k_y \,\mathcal{F}(k_x, k_y)$, and that $D_x = 0$\footnote{We leave an investigation of the \emph{minimal} symmetries needed to quantize the Berry curvature multipole moments for future work.}. 

First, note that $M_y$ directly implies $D_x = 0$, since $\mathcal{F}$ is odd under mirrors while $k_x$ is even under $M_y$, and therefore the integral of their product over the Brillouin zone vanishes. 
Also, $M_y$ implies that the portions of Berry curvature dipole moment in each of the two $K^2$ copies are identical, which implies that 
\begin{equation}
    D_y = 2D^{K^2}_{y}, \quad D^{K^2}_{y} \equiv \frac{1}{2\pi} \int_{-\pi}^{\pi}\int_{0}^{\pi}\mathrm{d}k_x\,\mathrm{d}k_y\, k_y \,\mathcal{F}(k_x, k_y).
\label{eq:quad_RP2}
\end{equation}

A crucial result, first shown in Ref.~\citenum{Chen2024}, is that the presence of $M_{y}$ quantizes the Chern number on $K^2$ to an integer, defined as 
\begin{equation}
    C^{K^2} 
    \equiv
    \frac{1}{2\pi}
    \int_{-\pi}^{\pi}\int_{0}^{\pi}\mathrm{d}k_x\,\mathrm{d}k_y\,\mathcal{F}(k_x, k_y).
    \label{eq:K2_chern}
\end{equation}
This quantization stems from the fact that the mirror symmetries cause the Berry curvature to vanish along the high-symmetry line $k_y = 0$. 
Thus, as explained in Ref.~\citenum{Chen2024}, we can treat the $K^2$ boundary as a single point when integrating
the Berry curvature, and the integral is a topological invariant.

The next step is to relate the Berry curvature dipole moment on $K^2$ to $C^{K^2}$.

We first change integration variables so that $(0,\pi/2)$ is our new origin, i.e., let $\bar{k}_y = k_y-\pi/2.$ We find
\begin{align}
    D^{K^2}_y
    &=
    \frac{1}{2\pi}\int_{-\pi}^{\pi} \int_{-\pi/2}^{\pi/2} \mathrm{d}\bar{k}_x\,\mathrm{d}\bar{k}_y\, (\bar{k}_y+\pi/2)\mathcal{F}(k_x,\bar{k}_y+\pi/2)
    \nonumber\\
    &=
    \frac{1}{2\pi}\int_{-\pi}^{\pi} \int_{-\pi/2}^{\pi/2} \mathrm{d}\bar{k}_x\,\mathrm{d}\bar{k}_y\,  \bar{k}_y\mathcal{F}(k_x,\bar{k}_y+\pi/2)+\frac{\pi}{2}C^{K^2}
    \nonumber\\
    &=
    \frac{\pi}{2}C^{K^2}.
\end{align}

In the last equality we used the fact that the integral vanishes since the integration region and distribution $\mathcal{F}$ are inversion-symmetric around $(\pi/2,\pi/2)$, which is a consequence of composing $\mathcal{G}_x$ and $M_y$.
Therefore,
\begin{equation}
    D_{y} = \pi C^{K^2}, \quad C^{K^2} \in \mathbb{Z},
\end{equation}
which explicitly indicates the quantization of $D_{y}$.

\subsection{Quadrupole moment}
In this section, we consider systems with three momentum-space nonsymmorphic symmetries:

\begin{equation}
\label{eq:BC_quadrupole_symms}
    \begin{split}
        &\overline{\mathcal{G}}_x: H(k_x, k_y) = U_1^{\dag}H(-k_x+\pi, k_y + \pi)U_1, \\
        &\overline{\mathcal{G}}_y: H(k_x, k_y) = U_2^{\dag}H(k_x + \pi, -k_y + \pi)U_2, \\
        &\mathcal{S}: H(k_x, k_y) = U_3^{\dag}H(k_y, -k_x+\pi)U_3,
    \end{split}
\end{equation}

with $U_i$ unitaries.
We will show that this leads to the quantization of the off-diagonal components of the Berry curvature quadrupole moment, $Q_{xy} = \frac{1}{\pi} \int_{\text{BZ}} \mathrm{d}^2\mathbf{k}\, k_x \, k_y \,\mathcal{F}(k_x, k_y)$.
In addition, these symmetries also force the diagonal components to vanish, \ie $Q_{xx}=Q_{yy} = 0$. 

Let us start by analyzing the composition of such symmetries.
Squaring $\overline{\mathcal{S}}$ yields an inversion-like symmetry about the $RP^2$ center, which we denote as $\overline{\mathcal{I}}$.
Composing $\overline{\mathcal{I}}$ with $\overline{\mathcal{G}_{x,y}}$ yields mirror symmetries $M_{y,x}$ about the origin in momentum space.
This directly implies that $Q_{xx} = Q_{yy} = 0$, since $\mathcal{F}$ is odd under mirrors while $k^2_i$ is even, and therefore the integral of their product over the Brillouin zone vanishes.

To show that $Q_{xy}$ is quantized, let us first note that, due to $M_x$ and $M_y$, the portions of Berry curvature quadrupole moment in each of the four $RP^2$ copies are identical, which implies that 
\begin{equation}
    Q_{xy} = 4Q^{RP^2}_{xy}, \quad Q^{RP^2}_{xy} \equiv \frac{1}{\pi} \int_{0}^{\pi}\int_{0}^{\pi}\mathrm{d}k_x\,\mathrm{d}k_y\, k_x \, k_y \,\mathcal{F}(k_x, k_y).
\label{eq:quad_RP2}
\end{equation}

We can once again use the results from Ref.~\citenum{Chen2024}, which states that the presence of $M_{x}$ and $M_{y}$ quantizes the Chern number on $RP^2$ to an integer, defined as 
\begin{equation}
    C^{RP^2} 
    \equiv
    \frac{1}{2\pi}
    \int_{0}^{\pi}\int_{0}^{\pi}\mathrm{d}k_x\,\mathrm{d}k_y\,\mathcal{F}(k_x, k_y).
    \label{eq:RP2_chern}
\end{equation}

\begin{figure}[t]
    \centering
    \includegraphics[scale=1]
    {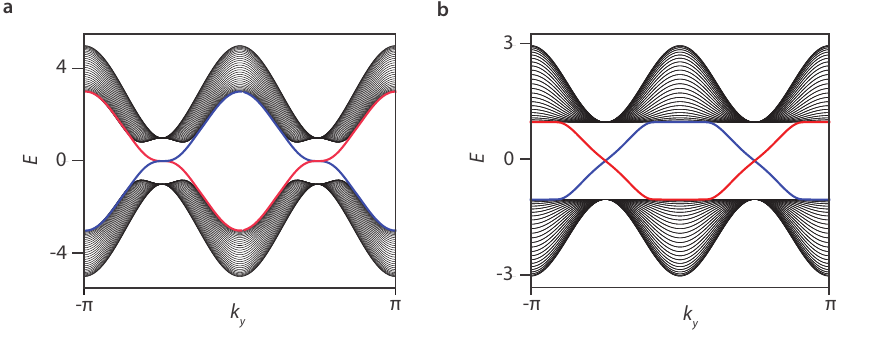}
    \caption{%
    (a) Energy dispersion for the dipolar Chern insulator (Eq.(8) of the main text) with $M = 2$ and open boundary conditions along $x$ (b) Energy dispersion for the quadrupolar Chern insulator (Eq.(11) of the main text) with $M = 1$ and open boundary conditions along $x$ .
    }
    \label{fig:sm_figure1}
\end{figure}

The next step is to relate both the Berry curvature dipole and quadrupole moments on the $RP^2$ to $C^{RP^2}$.
We start with the $RP^2$ dipole moment:
\begin{equation}
    D^{RP^2}_{i} \equiv \frac{1}{2\pi} \int_{0}^{\pi}\int_{0}^{\pi}\mathrm{d}k_x\,\mathrm{d}k_y\, k_i \,\mathcal{F}(k_x, k_y).
\end{equation}

We first change integration variables so that $(\pi/2,\pi/2)$ is our new origin, i.e., let $\bar{k}_x= k_x-\pi/2, \bar{k}_y = k_y-\pi/2.$ We find
\begin{align}
    D^{RP^2}_i
    &=
    \frac{1}{2\pi}\int_{-\pi/2}^{\pi/2} \int_{-\pi/2}^{\pi/2} \mathrm{d}\bar{k}_x\,\mathrm{d}\bar{k}_y\, (\bar{k}_i+\pi/2)\mathcal{F}(\bar{k}_x+\pi/2,\bar{k}_y+\pi/2)
    \nonumber\\
    &=
    \frac{1}{2\pi}\int_{-\pi/2}^{\pi/2} \int_{-\pi/2}^{\pi/2} \mathrm{d}\bar{k}_x\,\mathrm{d}\bar{k}_y\,  \bar{k}_i\mathcal{F}(\bar{k}_x+\pi/2,\bar{k}_y+\pi/2)+\frac{\pi}{2}C^{RP^2}
    \nonumber\\
    &=
    \frac{\pi}{2}C^{RP^2}.
\end{align}

In the last equality we used the fact that the integral vanishes since the integration region and distribution $\mathcal{F}$ are inversion-symmetric around $(\pi/2,\pi/2)$ due to $\overline{\mathcal{I}}$, while $\bar{k}_i$ is odd. 

Similarly, for the quadrupole moment:
\begin{align}
    Q^{RP^2}_{xy}
    &=
    \frac{1}{\pi}\int_{-\pi/2}^{\pi/2} \int_{-\pi/2}^{\pi/2} \mathrm{d}\bar{k}_x\,\mathrm{d}\bar{k}_y\,  (\bar{k}_x+\pi/2) (\bar{k}_y+\pi/2)\mathcal{F}(\bar{k}_x+\pi/2,\bar{k}_y+\pi/2)
    \nonumber\\
    &=
    \frac{1}{\pi}\int_{-\pi/2}^{\pi/2} \int_{-\pi/2}^{\pi/2} \mathrm{d}\bar{k}_x\,\mathrm{d}\bar{k}_y\,  \left(\bar{k}_x \bar{k}_y+\tfrac{\pi}{2}(\bar{k}_x+\bar{k_y})+\tfrac{\pi^2}{4}\right)\mathcal{F}(\bar{k}_x+\pi/2,\bar{k}_y+\pi/2)
    \nonumber\\
    &=
    \overline{Q}^{RP^2}_{xy}+\tfrac{2\pi^2}{4} C^{RP^2},
\end{align}
where $\overline{Q}^{RP^2}_{xy}$ is the quadrupole moment computed from the shifted origin $(\pi/2,\pi/2),$ and we have used the fact that the shifted dipole integrals vanish, as mentioned above.
The last step is to recognize that $\overline{Q}^{RP^2}_{xy} = 0$ due to $\mathcal{S}$, since the Berry curvature is even under this symmetry, while $\bar{k}_x \bar{k}_y$ is odd.
Finally, using~\cref{eq:quad_RP2}, we get
\begin{equation}
    Q_{xy} = 2\pi^2 C^{RP^2}, \quad C^{RP^2} \in \mathbb{Z},
\end{equation}
which explicitly indicates the quantization of $Q_{xy}$.

\section{Edge states}
Here, we discuss the bulk-boundary correspondence of the dipolar and quadrupolar Chern insulators introduced in the main text. 
These phases exhibit a quantized Chern number on subsets of the Brillouin zone, as discussed above, and serve as examples of the so-called Chern dartboard insulators~\cite{Chen2024}. 
These phases are also connected to delicate topology~\cite{schindler2021noncompact, nelson2022delicate} and returning Thouless pumps~\cite{zhu2023z, alexandradinata2024quantization}. 
As a result, the dipolar and quadrupolar Chern insulators host edge states with both chiralities on each edge~\cite{Chen2024}, reflecting the fact that the Chern number is vanishing on the full Brillouin zone.
In Fig.~\ref{fig:sm_figure1} we plot the edge states for the dipolar Chern insulator and quadrupolar Chern insulator models studied in the main text.

\bibliography{main}